\documentclass[12pt]{article}

\usepackage{amsmath, amssymb}

\allowdisplaybreaks[1]

\usepackage{color}

\newcommand{\eqa}{\begin{eqnarray}}
\newcommand{\eeqa}{\end{eqnarray}}
\newcommand{\beq}{\begin{equation}}
\newcommand{\eeq}{\end{equation}}

\newtheorem{dfn}{Definition}[section]
\newtheorem{thm}[dfn]{Theorem}
\newtheorem{rmk}[dfn]{Remark}

\newtheorem{cor}[dfn]{Corollary}
\newtheorem{emp}[dfn]{Example}
\newtheorem{cnj}[dfn]{Conjecture}

\newcommand{\nn}{\nonumber}
\newcommand{\f}{\frac}
\newcommand{\pa}{\partial}

\newcommand{\e}{\epsilon}

\newcommand{\al}{\alpha}
\newcommand{\pal}{\partial}
\newcommand{\p}{\partial}
\newcommand{\HH}{{\mathcal{H}}}
\newcommand{\bt}{{\bold{t}}}

\newcommand{\CC}{\mathbb{C}}
\newcommand{\la}{\lambda}

\newcommand{\F}{{\mathcal{F}}}

\newenvironment{prf}{\noindent {\it Proof} \ }{\hfill $\Box$}

\newenvironment{prfn}[1]{\noindent {\it Proof of #1} \ }{\hfill $\Box$}

\begin{document}
\title{On a class of integrable deformations of the integrable hierarchy of topological type associated to a semisimple Frobenius manifold}
\author{
{Si-Qi Liu$\,^a$, Paolo Rossi$\,^b$, Di Yang$\,^c$, Youjin Zhang$\,^a$}\\
{\small ${}^{a}$ Department of Mathematical Sciences, Tsinghua University,}\\
{\small Beijing 100084, P.R. China}\\
{\small ${}^{b}$ Dipartimento di Matematica ``Tullio Levi-Civita", 
Universit\'a degli Studi di Padova,}\\
{\small Via Trieste 63, 35121 Padova, Italy}\\
{\small ${}^{c}$ School of Mathematical Sciences, University of Science and Technology of China,}\\
{\small Hefei 230026, P.R. China}
}
\renewcommand{\thefootnote}{}
\footnotetext{E-mails:  \quad liusq@tsinghua.edu.cn,  \quad paolo.rossi@math.unipd.it,  \quad diyang@ustc.edu.cn,  \\ 
   youjin@tsinghua.edu.cn}
\renewcommand{\thefootnote}{\arabic{footnote}} 
\date{}
\maketitle
\begin{abstract}
Given a semisimple Frobenius manifold,
we construct a class of integrable 
deformations of its hierarchy of topological type. We show that
these integrable deformations have polynomial tau-structures, and   
conjecture that for the one-dimensional Frobenius manifold 
they give a universal object for integrable deformations of the Riemann--Hopf hierarchy having a tau-structure.
\end{abstract}

\section{Introduction}\label{Sect1}
The concept of Frobenius manifold was introduced by Dubrovin~\cite{Du0} (cf.~\cite{Du1}) to give a coordinate-free 
description of 2D topological field theories. It soon became one of the central subjects in 
Gromov--Witten theory, quantum singularity theory, and integrable systems.  

Recall that a \emph{Frobenius algebra} is a triple $(A, \,\langle\,,\rangle, e)$, where $A$ is a commutative associative algebra over $\mathbb{C}$ with unity~$e$ 
and $\langle\,,\rangle$ is a non-degenerate symmetric bilinear product, such that
$$\langle a_1 \cdot  a_2,a_3\rangle =\,\langle a_1,a_2 \cdot a_3\rangle, \quad \forall\,a_1,a_2,a_3\in A.$$
A \emph{Frobenius structure of charge~$d$} on an $n$-dimensional complex manifold~$M$ is a 
family of Frobenius algebras $(A_p=T_pM,\,\langle\,,\rangle_p, e_p)$,  $p\in M$, on the tangent spaces depending analytically on $p$ and satisfying the following axioms:
\begin{itemize}
\item[\textbf{FM1}]
The metric $\langle\,,\rangle$ on $M$ is flat. Denote by $\nabla$ the Levi-Civita connection of $\langle\,,\rangle.$ Then $\nabla \, e=0$.
\item[\textbf{FM2}]
Let $c$ denote the symmetric trilinear form on $TM$ defined by $c(x,y,z):=(x\cdot  y,z)$. 
The quadrilinear form $ (\nabla_w \, c) (x,y,z)$ must be symmetric. Here $x,y,z,w\in TM$.
\item[\textbf{FM3}]
A vector filed $E$ must be assigned on $M$ satisfying $\nabla\nabla E=0$, such that 
$$[E, x\cdot y]-[E,x]\cdot y-x\cdot [E,y]=x\cdot y,$$
and 
$$E\langle x,y\rangle-\langle[E,x],y\rangle-\langle x,[E,y]\rangle =(2-d)\langle x,y\rangle.$$
\end{itemize}
A complex manifold endowed with a Frobenius structure of charge~$d$ is called a \emph{Frobenius manifold of charge~$d$}. 
The vector field $E$ is called the {\it Euler vector field}. 

Let $M$ be an $n$-dimensional Frobenius manifold.
Since $\langle\,,\rangle$ is a flat metric, we can choose a system of flat coordinates 
$(v^1,\dots,v^n)={\bf v}$ such that the matrix $(\eta_{\alpha\beta})$ defined by
$\eta_{\alpha\beta}=\langle\frac{\p}{\p v^\alpha},\frac{\p}{\p v^\beta}\rangle$
is constant, symmetric and invertible. 
Here and below, we use $(\eta_{\alpha\beta})$ or 
its inverse $(\eta^{\alpha\beta}):=(\eta_{\alpha\beta})^{-1}$ to lower or raise the Greek indices, 
free Greek indices take integer values from~1 to~$n$, the 
Einstein summation convention is assumed for repeated upper and lower Greek indices.
Since $\nabla e=0$ we can further choose the flat coordinates so that 
\beq\label{unitypv}
e=\frac{\p}{\p v^1}.
\eeq 

In~\cite{Du1} Dubrovin defined the 
principal hierarchy of~$M$ (see~\eqref{PHeq} below for detail), which is 
an integrable hierarchy of evolutionary PDEs for~${\bf v}$ of the form
\beq\label{ph-intro}
\frac{\p v_\alpha}{\p t^{\beta,m}}=\frac{\p\Omega^{[0]}_{\alpha,0;\beta,m}({\bf v})}{\p x},\quad m\geq 0,
\eeq
where $v_\alpha=\eta_{\alpha\beta}v^\beta$, and $\Omega^{[0]}_{\alpha,0;\beta,m}$ 
are certain holomorphic functions  
satisfying $\Omega^{[0]}_{\alpha,0;1,0}({\bf v})=v_\alpha$ (see Section~\ref{section2} for detail).
Dubrovin also~\cite{Du1} constructed the topological solution ${\bf v}_{\rm top}(\bt)$ to the principal hierarchy
 and the genus~0 topological free energy $\F_0(\bt)$ (cf.~\cite{DW, DZ-norm}). 
Here 
$\bt=(t^{\alpha,m})_{\al=1,\dots,n,\,m\ge0}$ is an infinite vector often called {\it times} or {\it coupling constants}. 
It was shown in~\cite{DZ-vira} (cf.~\cite{DZ-norm,LT}) that~$\mathcal{F}_0(\bt)$ satisfies the genus~0 Virasoro constraints:
\beq
e^{-\e^{-2}\mathcal{F}_0(\bt)} L_i \bigl(e^{\e^{-2}\mathcal{F}_0(\bt)}\bigr) = O(1), \quad \epsilon \to 0,
\eeq
where the $L_i$, $i\ge-1$, are the Virasoro operators of~$M$~\cite{DZ-vira, DZ-norm, LT}, which have the form
\begin{align}
&\e^2 \sum_{m,m'} a^{\al,m;\beta,m'} \frac{\p^2}{\p t^{\al,m}\p t^{\beta,m'}}+\sum_{m,m'} b^{\al,m}_{\beta,m'}\, \tilde t^{\beta,m'}\frac{\p }{\p t^{\al,m}}+\frac{1}{\e^2}\sum_{m,m'} c_{\al,m;\beta,m'}\tilde t^{\al,m} \tilde t^{\beta,m'}, \label{def-L}
\end{align}
with $a^{\al,m;\beta,m'},\,b^{\al,m}_{\beta,m'},\,c_{\al,m;\beta,m'}$ being constants independent of~$\e$,  and
$\tilde t^{\alpha,m}=t^{\alpha,m}-\delta^{\alpha,1} \delta^{m,1}$.
Here and in what follows $\sum_m:=\sum_{m\ge0}$.

Consider a family of evolutionary PDEs of divergence form for ${\bf w}=(w_\al)_{\al=1}^n$
\beq\label{pert-int-intro}
\frac{\p w_\alpha}{\p t^{\beta,m}}=\frac{\p \Omega_{\alpha,0;\beta,m}({\bf w}_0, {\bf w}_1, \dots;\e)}{\p x},\quad m\geq 0,
\eeq
with ${\bf w}_0={\bf w}$, $\frac{\p}{\p t^{1,0}}=\frac{\p}{\p x}$ 
and $\Omega_{\alpha,0;\beta,m}({\bf w}, {\bf w}_1, \dots;\e)$ being power series of the form
\beq
\Omega_{\alpha,0;\beta,m}({\bf w}, {\bf w}_1, \dots;\e)=\Omega^{[0]}_{\alpha,0;\beta,m}({\bf w})
+\sum_{g\geq 1} \e^{2g} \Omega^{[g]}_{\alpha,0;\beta,m}({\bf w}, {\bf w}_1, \dots), \quad m\ge0,
\eeq
where $\Omega^{[0]}_{\alpha,0;\beta,m}$ are the functions in~\eqref{ph-intro}, and 
$\Omega^{[g]}_{\alpha,0;\beta,m}({\bf w}, {\bf w}_1, \dots)\in \mathcal{A}_{\bf w}^{[2g]}$, $g\geq 1$.
Here, $\p/\p x$ can be understood as $\sum_{k\ge0} w_{\alpha,k+1} \frac{\p }{\p w_{\alpha,k}}$, and 
 $\mathcal{A}_{\bf w}^{[m]}$ denotes the set of elements in 
$\mathcal{O}(M)[w_{\alpha,1},w_{\alpha,2},\dots|\alpha=1,\dots,n]$
that are homogeneous of degree~$m$ with respect to $\deg w_{\alpha,k}:=k$ ($k\ge0$).
We call such a family~\eqref{pert-int-intro}
 an {\it integrable deformation of the principal hierarchy~\eqref{ph-intro} of divergence form}\footnote{For simplicity we do not consider odd powers of~$\e$ here.} if equations in~\eqref{pert-int-intro} pairwise commute:
\beq\label{flowscomm}
\frac{\p}{\p t^{\gamma,m'}}\frac{\p w_\alpha}{\p t^{\beta,m}}=\frac{\p}{\p t^{\beta,m}}\frac{\p w_\alpha}{\p t^{\gamma,m'}},\quad \forall\,m,m'\geq 0.
\eeq
We note that the above set-up forces $\Omega_{\alpha,0;1,0}({\bf w})=w_\alpha$.

\begin{dfn}\label{def-tau-symm}
An integrable deformation~\eqref{pert-int-intro} of the principal 
hierarchy of divergence form
is said to possess a {\it tau-structure} if the following two conditions are satisfied:
\begin{itemize}
\item[a)] There exist $\widehat \Omega_{\beta,m;\gamma,m'} 
= \sum_{g\geq 0} \e^{2g} \widehat \Omega^{[g]}_{\beta,m;\gamma,m'}$ with
$\widehat \Omega^{[g]}_{\beta,m;\gamma,m'}\in\mathcal{A}_{\bf w}^{[2g]}$, such that 
\beq\label{propertya}
\frac{\p \Omega_{1,0;\beta,m}}{\p t^{\gamma,m'}}=\frac{\p \widehat \Omega_{\beta,m;\gamma,m'}}{\p x},\quad m,m'\ge0.
\eeq
\item[b)] We have, in particular,
\beq \label{propertyb}
\Omega_{1,0;\alpha,0}=w_\alpha.
\eeq
\end{itemize}
\end{dfn}
We note that an integrable deformation~\eqref{pert-int-intro} of the principal 
hierarchy of divergence form having a tau-structure does not necessarily have a polynomial hamiltonian structure. 

The goal of this paper is to introduce for a semisimple Frobenius manifold a class of integrable deformations of the principal 
hierarchy of divergence form having a tau-structure; 
for the definition of semisimplicity see e.g. \cite{Du1} or Section~\ref{section2}.

Let $M$ be a semisimple Frobenius manifold. We first recall the following theorem.

\smallskip

\noindent {\bf Theorem~A} (\cite{DZ-norm}).  {\it 
Write
\beq\label{genusF}
\F=\F(\bt;\e) = \sum_{g\ge0} \e^{2g-2} \F_g(\bt).
\eeq
Then there exist functions $F_g({\bf v}_0={\bf v}, {\bf v}_1, \dots, {\bf v}_{3g-2})$, $g\ge1$, such that 
\beq
\F_g(\bt) :=
F_g\biggl({\bf v}_{\rm top}(\bt), \frac{\p {\bf v}_{\rm top}(\bt)}{\p x},\dots,\frac{\p^{3g-2} {\bf v}_{\rm top}(\bt)}{\p x^{3g-2}}\biggr), \quad g\ge1,
\eeq
satisfy
\begin{align}
& L_i \bigl(e^{\F(\bt;\e)}\bigr) = 0, \quad i\ge-1, \label{Viraconstraints}\\
&\sum_{m} \tilde t^{\alpha,m} \frac{\p e^{\F(\bt;\e)} }{\p t^{\alpha,m}}  + \e \frac{\p e^{\F(\bt;\e)} }{\p \e} + \frac{n}{24} e^{\F(\bt;\e)} = 0.
\end{align}
Moreover, the functions $F_g$, $g\ge1$, are uniquely determined up to an additive constant in $F_1$.
Furthermore, $F_1$ can be given by the formula
\[F_1({\bf v}, {\bf v}_1)=\frac1{24}\log\det(c^\al_{\beta\gamma}({\bf v}) {\bf v}^\gamma_1)+G({\bf v}),\]
with $c^\al_{\beta\gamma}({\bf v})$ being structure constants and $G({\bf v})$ being G-function (see \cite{Du2, DZ-oneloop, DZ-norm, Getzler});  
for $g\ge2$, $F_g$
depends polynomially on $v^\gamma_k$ ($k\ge 2$) and rationally on $v^\gamma_1$, and satisfies
\[\deg F_g=2g-2,\quad \overline{\deg}\, F_g\le 3g-3,\]
with the degrees defined via $\deg v^\alpha_k:=k$ ($k\ge0$), $\overline{\deg} v^\alpha_k:=k-1$ ($k\ge1$), $\overline{\deg} v^\alpha_0=0$.}
\smallskip

Now recall that the {\it integrable hierarchy of topological type associated to~$M$}, also 
well known as the {\it Dubrovin--Zhang hierarchy of~$M$}, is defined~\cite{DZ-norm} 
as the family of evolutionary PDEs obtained after substitution of the following quasi-triviality map 
\beq\label{quasitrivialmap}
u_\alpha = v_\alpha + \sum_{g\ge1} \e^{2g} \frac{\p^2 F_g({\bf v}_0, {\bf v}_1, \dots, {\bf v}_{3g-2})}{\p t^{\alpha,0} \p x}
\eeq
into the principal hierarchy~\eqref{ph-intro}. It has the form
\beq\label{DZhier}
\frac{\p u_\alpha}{\p t^{\beta,m}} = \frac{\p \Omega^{\rm DZ}_{\alpha,0;\beta,m}}{\p x}, \quad m\ge0.
\eeq
Here, 
$\Omega^{\rm DZ}_{\alpha,0;\beta,m}$ are power series of~$\e^2$. 
By definition and by the results in~\cite{BPS1,BPS2}, the hierarchy
\eqref{DZhier} is an integrable deformation of the principal 
hierarchy of divergence form having a tau-structure, and moreover,  the function $Z$
is a special tau-function of~\eqref{DZhier} (cf.~\cite{DZ-norm, Witten}), called the topological tau-function of~\eqref{DZhier} 
or the topological partition function of~$M$. 
Below, unless explicitly indicated, we always assume that $M$ is semisimple.

Let $L$ be a linear operator of the form~\eqref{def-L}.
Similarly to~\cite{FP, G1}, we consider the following initial value problem:
\begin{eqnarray}
\frac{\p Z_L}{\p s}&=&L (Z_L),\label{def-zqa}\\
Z_L|_{s=0}&=&Z.\label{def-zqb}
\end{eqnarray}
The unique solution $Z_L=Z_L(\bt;s;\e)$ to~\eqref{def-zqa}--\eqref{def-zqb} is called the {\it $L$-deformed partition function}.  
Obviously, $Z_L(\bt;s;\e)=e^{s L} Z(\bt;\e)$.
\begin{thm}\label{H}
If the linear operator $L$ satisfies
\beq
\label{zero-const-L}
e^{-\e^{-2}\F_0(\bt)}  L \bigl(e^{\e^{-2}\F_0(\bt)}\bigr) = O(1) ,\quad \e\to 0,
\eeq
then the $L$-deformed partition function $Z_L$ has the following genus expansion:
\begin{equation}
Z_L=\exp\biggl( \, \sum_{g\geq0} \e^{2g-2} \HH_g(\bt;s)\biggr),\label{genus-exp}
\end{equation}
where 
$\HH_0(\bt;s) \equiv \mathcal{F}_0(\bt)$, and 
for $g\geq 1$ there exists $H_g=H_g({\bf v}_0, \dots, {\bf v}_{3g-2};s)$ such that 
\beq
 \HH_g(\bt;s)=H_g\biggl({\bf v}(\bt),\frac{\p {\bf v}(\bt)}{\p x},\dots,\frac{\p^{3g-2} {\bf v}(\bt)}{\p x^{3g-2}};s\biggr).
\eeq 
Moreover, 
for $g=1$, $H_1-F_1$ does not depend on $v^\gamma_k$ ($k\ge1$); 
for $g\ge2$, $H_g$
depends polynomially on $v^\gamma_k$ ($k\ge 2$) and rationally on $v^\gamma_1$, and satisfies
\[\deg H_g=2g-2,\quad \overline{\deg}\,H_g\le 3g-3.\]
\end{thm}
The proof is in Section~\ref{section2}.

Let $L$ be a linear operator of the form~\eqref{def-L} satisfying~\eqref{zero-const-L}.
The {\it $L$-deformed hierarchy} is defined as the substitution 
of the following quasi-Miura map 
\begin{equation}\label{quasi-Miura}
w_\alpha=v_\alpha+\p_x\p_{t^{\alpha,0}} \Bigl(\,\sum_{g\geq 1} \e^{2g} H_g\Bigr)
\end{equation}
into the principal hierarchy~\eqref{ph-intro}. 

Using the method given in~\cite{BPS1, BPS2} (cf.~\cite{Buryak}), we will prove in Section~\ref{section4} the following theorem,  
which is a joint work with Boris Dubrovin.
\begin{thm}\label{ab} 
Let $L$ be an operator of the form~\eqref{def-L} satisfying~\eqref{zero-const-L}.
The $L$-deformed hierarchy of $M$ is an integrable deformation of the principal 
hierarchy of divergence form having a tau-structure.
\end{thm}

Recall that 
the Virasoro operators $L_i$, $i\ge-1$, of~$M$ satisfy the following commutation relations:
\beq
[L_i, L_j] = (i-j) L_{i+j}, \quad i,j\ge-1.
\eeq
We call ${\rm Vira}:=\oplus_{i=-1}^\infty \CC L_i$ the Virasoro algebra of~$M$. 
A deformation of ${\rm Vira}$, denoted by ${\rm Vira}_{\rm like}$ and called the {\it Virasoro-like algebra of~$M$}, 
is an infinite-dimensional Lie algebra constructed in~\cite{LYZZ}:
\beq
{\rm Vira}_{\rm like} = \CC {\rm id} \oplus {\rm Span}_{\CC}\{L_{i,2j} \mid i\ge-1, \, 0\le j \le [(i+1)/2]\} .
\eeq
For the definition of the operators $L_{i,2j}$ (called Virasoro-like operators), see~\cite{LYZZ} or Section~\ref{section5}.
The infinite-dimensional Lie algebra ${\rm Vira}_{\rm like}$
contains three interesting Lie subalgebras~\cite{LYZZ}: the first one is ${\rm Vira}$, 
the second one is ${\rm Vira}(\frac12)$ (see~\cite{LYZZ}), and
 the third one is the commutative Lie algebra ${\rm Span}_{\CC}\{L_{2j-1,2j}\mid j\ge1\}$ whose significance 
 has been clarified in~\cite{FP} (cf.~\cite{G1}). These considerations do not require semisimplicity~\cite{LYZZ}.

The following theorem was proved in~\cite{LYZZ} without the semisimplicity assumption.

\noindent {\bf Theorem~B} (\cite{LYZZ}). {\it 
For $i\ge-1$ and $0\leq j \leq [(i+1)/2]$, we have}
\begin{equation}
e^{-\e^{-2} \F_0(\bt)} L_{i,2j} \bigl(e^{\e^{-2}\F_0(\bt)}\bigr) = O(1),\qquad \epsilon\rightarrow 0.
\end{equation}

We note that the constraints given by $L_{1,2}$ and $L_{2,2}$ appeared in~\cite{EHX, LT}, 
and the constraints given by 
$L_{2k-1,2k}$, $k\ge1$, were obtained in~\cite{FP} (see also~\cite{G1}).

Define 
\beq
Z_{\rm like} = Z_{\rm like} (\bt; {\bf r};\e) := \prod_{m=1}^\infty \prod_{k=1}^{[(m+1)/2]} e^{r_{m, 2k} L_{m,2k}} ( Z ),
\eeq
where ${\bf r}=(r_{1,2},r_{2,2},r_{3,2},\dots)$, 
the product means operator compositions, 
and we fix a particular ordering of the compositions. 
The {\it Virasoro-like hierarchy associated to $Z_{\rm like}$ (via successive applications of Virasoro-like operators)} is defined as the substitution of the quasi-Miura transformation
\beq
w_\alpha= \e^2 \frac{\p^2 \log Z_{\rm like} (\bt; {\bf r};\e) }{\p t^{\alpha,0} \p x} 
\eeq
in the principal hierarchy. 

The following result, whose discovery was joint with Boris Dubrovin, 
 was announced in~\cite{LYZZ}, and we will give its proof in Section~\ref{section4}.

\begin{thm} \label{B-o}
The Virasoro-like hierarchy associated to $Z_{\rm like}$ is an integrable deformation of the principal 
hierarchy of divergence form having a tau-structure.
\end{thm}

\noindent {\bf Organization of the paper.}
In Section~\ref{section2}, we give a brief review of Frobenius manifold. 
In Section~\ref{section3}, we prove Theorem~\ref{H}.
In Section~\ref{section4}, we prove Theorem~\ref{ab}.
In Section~\ref{section5}, we give more details about the construction of the Virasoro-like hierarchy and give examples for the one-dimensional Frobenius manifold.

\noindent {\bf Acknowledgements.}
We wish to thank Boris Dubrovin for several very helpful discussions. 
The work is partially supported by NSFC No.~12371254, No.~12571266.
Part of the work of D.~Y. was carried out while he was a postdoc at SISSA and MPIM; 
he thanks both institutions for excellent working conditions.
The work of P.~R. is supported by the University of Padova and is affiliated to the INFN under the national project MMNLP and to the INdAM group GNSAGA. Part of his work was carried out during a visit to the MPIM at Bonn.

\section{Review of the theory of Frobenius manifolds}\label{section2}
In this section, we give a review of 
basic terminologies about 
Frobenius manifolds.

Let $M$ be a Frobenius manifold.  As in Section~\ref{Sect1}, take ${\bf v}=(v^1,\dots,v^n)$ a system of flat coordinates satisfying~\eqref{unitypv}. 
Let 
$$c_{\alpha\beta\gamma}({\bf v})
:=\langle\frac{\p}{\p v^\alpha}\cdot\frac{\p}{\p v^\beta},\frac{\p}{\p v^\gamma}\rangle.$$ 
By the axiom {\bf FM2} there locally exist a function $F=F({\bf v})$ such that 
$$c_{\alpha\beta\gamma}({\bf v})=\frac{\p^3 F({\bf v})}{\p v^\alpha\p v^\beta \p v^\gamma}.$$
The associativity of the multiplication written in terms of~$F$ reads as follows:
$$
\frac{\p^3 F}{\p v^\alpha\p v^\beta \p v^\rho}\eta^{\rho\sigma} \frac{\p^3 F}{\p v^\sigma\p v^\phi \p v^\psi}
=\frac{\p^3 F}{\p v^\psi\p v^\beta \p v^\rho}\eta^{\rho\sigma} \frac{\p^3 F}{\p v^\sigma\p v^\phi \p v^\alpha}.
$$
This set of equations are called the {\it WDVV associativity equations}, named after Witten--Dijkgraaf--Verlinde--Verlinde.

Recall that the \emph{Dubrovin connection} $\widetilde{\nabla}$ is an affine connection on $M\times\mathbb{C}^*$ defined by 
\begin{align}
&\widetilde{\nabla}_a b=\nabla_a b+z\,a\cdot b, \nn\\
&\widetilde{\nabla}_{\frac{d}{d z}} b=\pal_z b+ E\cdot b-\frac{1}{z} \mu \, b,\quad
\widetilde{\nabla}_{\frac{d}{d z}} \frac{d}{d z}=\widetilde{\nabla}_{b}{\frac{d}{d z}} =0\nn
\end{align}
for arbitrary horizontal holomorphic vector fields $a,b \in \textrm{Vect}(M\times\mathbb{C}^*)$.
Here $\mu:=1-\frac{d}{2}-\nabla E$. It was shown in~\cite{Du1} that the Dubrovin connection $\widetilde{\nabla}$ is flat.

For simplicity we assume from now on  that the Euler vector field~$E$ is diagonalizable, namely, in 
a suitable flat coordinate system satisfying~\eqref{unitypv} it has the expression
\begin{align}
&E=\sum_{\beta=1}^n \Bigl(\Bigl(1-\frac{d}2-\mu_\beta\Bigr) v^\beta + r^\beta\Bigr) \frac{\p}{\p v^\beta},   \label{E517} 
\end{align}
where $\mu_1,\dots,\mu_n, r^1,\dots,r^n$ are constants with 
$\mu_1=-\frac{d}2$ and $r^1=0$. We note that $\mu={\rm diag}(\mu_1,\dots,\mu_n)$.
The flatness of~$\tilde \nabla$ ensures~\cite{Du1,Du2,DZ-norm} the existence of analytic functions $\theta_{\alpha,m}({\bf v})$, $m\ge0$, satisfying  
\begin{align}
&
\theta_{\alpha,0}({\bf v})=v_\alpha, \label{theta000511}\\
&
\frac{\p^2 \theta_{\gamma,m+1}({\bf v})}{\p v^\alpha\p v^\beta} = c^\sigma_{\alpha\beta}({\bf v}) 
\frac{\p \theta_{\gamma,m}({\bf v})}{\p v^\sigma}, \quad m\ge0,\label{theta1c}\\
&\bigl\langle \nabla \theta_\alpha({\bf v};z),  \nabla \theta_\beta({\bf v};-z) \bigr\rangle = \eta_{\alpha\beta}, 
\label{orthogtheta511} \\
&
\frac{\p\theta_{\alpha,m+1}({\bf v})}{\p v^\iota} = \theta_{\alpha,m}({\bf v}), \quad  m\geq0, \label{thetacali512}\\
& E (\p_{\beta}(\theta_{\alpha,m}({\bf v}))) = 
(m+\mu_\alpha+\mu_\beta) \, \p_{\beta}(\theta_{\alpha,m}({\bf v})) + 
\sum_{k=1}^m (R_k)^\gamma_\alpha \, \p_{\beta} (\theta_{\gamma,m-k}({\bf v})) , \quad m\geq 0. \label{theta2c} 
\end{align}
Here, 
$$
\theta_\alpha({\bf v};z)= \sum_{m} \theta_{\al,m}({\bf v}) z^m, 
$$
and $R=R_1+R_2+\cdots$ is a constant matrix which together with $\mu$ form the monodromy data of $\widetilde \nabla$ around $z=0$.
Recall that $(\mu,R)$ satisfies the following relations:
\begin{equation*}
\eta_{\al\gamma}(R_k)^\gamma_\beta =(-1)^{k+1} \eta_{\beta\gamma}(R_k)^\gamma_{\alpha},\quad [\mu, R_k]=k R_k,\quad k\ge0.
\end{equation*}

The principal hierarchy associated to~$M$ \cite{Du1,Du2, DZ-norm} is defined by 
\beq\label{PHeq}
\f{\pa v^\al}{\pa t^{\beta,m}}=\eta^{\al\gamma}\pa_x \Bigl(\f{\pa\theta_{\beta,m+1}}{\pa v^\gamma}\Bigr)=\eta^{\delta \gamma}c^\al_{\delta\epsilon}\f{\pa\theta_{\beta,m}}{\pa v^\gamma}v^{\epsilon}_x, \quad m\ge0.
\eeq
Following~\cite{Du1,Du2, DZ-norm}, define  
the genus zero two-point correlation functions $\Omega_{\al,m_1;\beta,m_2}^{[0]}({\bf v})$ by means of generating series as follows:
\begin{equation}\label{two-point-genus-zero}
\sum_{m_1, m_2} \Omega^{[0]}_{\al,m_1;\beta,m_2}({\bf v}) z^{m_1} w^{m_2}
:=\frac{\langle\nabla\theta_\al({\bf v};z),\nabla\theta_\beta({\bf v};w)\rangle-\eta_{\al\beta}}{z+w}.
\end{equation}
They have the following properties:
\begin{align}
&\Omega^{[0]}_{\al,m_1;\beta,m_2}({\bf v}) =  \Omega^{[0]}_{\beta,m_2;\al,m_1}({\bf v}), \label{taustr0-1}\\
&\frac{\p \Omega^{[0]}_{\alpha,m_1;\beta,m_2}({\bf v}) }{\p t^{\gamma,m_3}}=\frac{\p\Omega^{[0]}_{\alpha,m_1;\gamma,m_3}({\bf v}) }{\p t^{\beta,m_2}}. \label{taustr0-2}
\end{align}

The topological solution ${\bf v}_{\rm top}(\bt)$ to the principal hierarchy \eqref{PHeq} can be defined as the unique solution to the 
genus~0 Euler--Lagrange equation~\cite{Du1,Du2,DZ-norm}: 
\begin{equation}
\label{Euler-Lagrange}
\sum_{m} \tilde t^{\al,m}\f{\pa \theta_{\al,m}}{\pa v^{\gamma}}(v_{\rm top}(\bt))=0.
\end{equation}
The genus~0 topological free energy~$\mathcal{F}_0$ is defined~\cite{Du1} by 
\beq
\mathcal{F}_0(\bt) = \sum_{m_1,m_2} \tilde t^{\al,m_1} \tilde t^{\beta,m_2} \Omega^{[0]}_{\al,m_1;\beta,m_2} ({\bf v}_{\rm top}(\bt)).
\eeq
It has the property
\beq
\frac{\pa^2\mathcal{F}_0(\bt)}{\p t^{\al,m_1} \p t^{\beta,m_2}} = \Omega^{[0]}_{\al,m_1;\beta,m_2} ({\bf v}_{\rm top}(\bt)), \quad m_1,m_2\ge0.
\eeq

Let $M$ be a Frobenius manifold. The intersection form of $M$ is defined by
$$(\omega_1,\omega_2)=i_E (\omega_1 \cdot \omega_2),\quad \forall\,\omega_1,\omega_2 \in Vect^*(M).$$ In the flat coordinate system, it has the expression
$$g^{\al\beta}=(dv^\al,dv^\beta)=E^\e c^{\alpha\beta}_\e.$$
The Frobenius structure implies that $\eta^{\al\beta}$ and $g^{\al\beta}$ form a flat pencil. 
The flat coordinates $p_\al({\bf v};\la)$ to the pencil $$g^{\al\beta}-\la \, \eta^{\al\beta}$$ are called {\it $\lambda$-periods} of $M$. 
Denote by $\phi_\alpha=\nabla p_\al$ the gradients of periods. 

Following~\cite{DZ-norm} (cf.~\cite{Du3,DZ-toda}), define
\begin{equation}\label{natural_with_nu}
\hat\phi_\al(\la,\nu):=\int_{0}^\infty \f{dz}{z^{1-\nu}}~e^{-\la z}\!\sum_{p\in
\mathbb{Z}+\f{1}{2}} a_{\beta,p} \big(z^{p+\mu}z^R\big)^\beta_\al,
\end{equation}
where $a_{\al,p},\,p\in \mathbb{Z}+\f{1}{2}$,
are creation and annihilation operators:
\begin{equation*}\label{creation_annihilation}
a_{\al,p}=\epsilon \f{\pa}{\pa
t^{\al,p-\f{1}{2}}}~(p>0),~a_{\al,p}=\f{1}{\epsilon}(-1)^{p+\f{1}{2}}\eta_{\al\beta}
\tilde t^{\beta,-p-\f{1}{2}}~(p<0).
\end{equation*}
Explicitly,
\begin{equation*}\label{natural_with_nu_explicit}
\hat\phi_\al(\la,\nu)=\sum_{p\in\mathbb{Z}+\f{1}{2}} a_{\beta,p} \sum_{r\geq 0} \bigl(\big[e^{R\pa_{\nu}}\big]_r \Gamma(\mu+p+r+\nu)\la^{-(\mu+p+r+\nu)}\la^{-R}\bigr)^\beta_\al.
\end{equation*}

The regularized stress-energy tensor, introduced in \cite{DZ-norm,DZ-toda}, is defined by 
\begin{equation*}\label{Virasoro_field_with_nu}
T(\la,\nu)=-\f{1}{2}:\pa_\la \hat \phi_\al(\la,\nu) G^{\al\beta}(\nu)\pa_\la \hat \phi_\beta(\la,-\nu):+\f{1}{4\la^2}{\rm tr}\Big(\f{1}{4}-\hat\mu^2\Big).
\end{equation*}
Here $$G^{\al\beta}(\nu):=-\f{1}{2\pi} \bigl[(e^{\pi i R}e^{\pi i (\mu+\nu)}+e^{-\pi i R}e^{-\pi i (\mu+\nu)})\eta^{-1}\bigr]^{\al\beta}$$ and $``:\,:"$ denotes the normal ordering (putting the annihilation operators on the right).
Following~\cite{DZ-norm} (cf.~\cite{DZ-toda, LYZZ}), introduce $L_i(\nu)$, $i\ge-1$, by  
\begin{equation}\label{Virasoro_field_with_nu_expand}
T(\la,\nu)=\sum_{i\in\mathbb{Z}}\f{L_i(\nu)}{\la^{i+2}}.
\end{equation}
Then the Virasoro operators $L_i$ in Section~\ref{Sect1} can be defined~\cite{DZ-norm, DZ-toda} (cf.~\cite{DZ-vira}) by
\beq
L_i:= \lim_{\nu\to 0} L_i(\nu), \quad i\ge -1.
\eeq

Finally we recall the definition of semisimplicity, which will be the main assumption for the later sections. A nonzero element~$a$ of 
a Frobenius algebra~$A$ is called a {\it nilpotent} if $a^k=0$ for some $k\ge1$. 
A Frobenius algebra is called {\it semisimple} if it contains no nilpotent.
A point $p$ of a Frobenius manifold~$M$ is called a {\it semisimple point} if $T_pM$ is semisimple. 
A Frobenius manifold is called {\it semisimple} if its generic points are semisimple.

\section{The $L$-deformed partition function}\label{section3}

In this section, we prove Theorem~\ref{H}.

\begin{prfn}{Theorem~\ref{H}} 
It is obvious to see that equations \eqref{def-zqa}, \eqref{def-zqb} are equivalent to 
\begin{align}
&\frac{\p e^{\HH(\bt;s;\e)}}{\p s}=L\bigl( e^{\HH(\bt;s;\e)} \bigr),\label{der-e-H}\\
&\HH(\bt;s=0;\e)=\F(\bt;\e). \label{ini-H}
\end{align}
Consider 
\beq \label{genus-expan-H}
\HH=\HH(\bt;s;\e) =:\sum_{g\geq 0} \e^{2g-2} \HH_g(\bt;s).
\eeq 
We will show that equations \eqref{der-e-H}--\eqref{genus-expan-H} have a unique solution $\HH_g=\HH_g(\bt;s)$, $g\ge0$.

Note that equation~\eqref{der-e-H} is equivalent to 
\begin{align}
\frac{\p \HH}{\p s}=\,&\e^2\sum_{m,n\geq 0} a^{\al,m;\beta,n} \biggl(\frac{\p \HH}{\p t^{\al,m}}\frac{\p\HH}{\p t^{\beta,n}}+\frac{\p^2 \HH}{\p t^{\al,m}\p t^{\beta,n}}\biggr)+\sum_{m,n\geq 0} b^{\al,m}_{\beta,n} \tilde t^{\beta,n}\frac{\p \HH }{\p t^{\al,m}}\nn\\
&+\frac{1}{\e^2}\sum_{m,n\geq 0} c_{\al,m;\beta,n}\tilde t^{\al,m}\tilde t^{\beta,n}. \label{HHs}
\end{align}
Then by comparing coefficients of powers of~$\e$ we know that \eqref{der-e-H}, \eqref{genus-expan-H} are equivalent to the following equations:
\begin{align}
\frac{\p \HH_0}{\p s}=\,&\sum_{m,n\geq 0} a^{\al,m;\beta,n} \frac{\p \HH_0}{\p t^{\al,m}}\frac{\p\HH_0}{\p t^{\beta,n}}+\sum_{m,n\geq 0} b^{\al,m}_{\beta,n} \tilde t^{\beta,n}\frac{\p \HH_0 }{\p t^{\al,m}}\nn\\
&+\sum_{m,n\geq 0} c_{\al,m;\beta,n}\tilde t^{\al,m}\tilde t^{\beta,n},\label{HH0s}\\
\frac{\p \HH_g}{\p s}=\,&\sum_{m,n\geq 0} a^{\al,m;\beta,n} \biggl(\,\sum_{k=0}^g \frac{\p \HH_k}{\p t^{\al,m}}\frac{\p\HH_{g-k}}{\p t^{\beta,n}}+\frac{\p^2 \HH_{g-1}}{\p t^{\al,m}\p t^{\beta,n}}\biggr)\nn\\
&+\sum_{m,n\geq 0} b^{\al,m}_{\beta,n} \tilde t^{\beta,n}\frac{\p \HH_g }{\p t^{\al,m}},\quad g\geq 1. \label{HHgs}
\end{align}
The condition~\eqref{ini-H} can be equivalently written as
\beq\label{Hinig}
\HH_g(\bt;0)=\F_g(\bt),\quad g\ge0.
\eeq

Noticing that the condition~\eqref{zero-const-L} is equivalent to
\begin{equation}\label{genus0-F0}
\sum_{m,m'} a^{\al,m;\beta,m'}\frac{\p \F_0}{\p t^{\al,m}}\frac{\p \F_0}{\p t^{\beta,m'}}+\sum_{m,m'} b^{\al,m}_{\beta,m'} \tilde t^{\beta,m'}\frac{\p \F_0}{\p t^{\al,m}}+\sum_{m,m'} c_{\al,m;\beta,m'}\tilde t^{\al,m}\tilde t^{\beta,m'}=0,
\end{equation}
we find that solution $\HH_0(\bt;s)$ to~\eqref{HH0s} and the $g=0$ equation of~\eqref{Hinig} exists and must satisfy that
 $$\HH_0(\bt;s)\equiv\F_0(\bt).$$

Denote 
\beq\label{D-op}
D:=\sum_{m,m'} a^{\al,m;\beta,m'} \biggl(\frac{\p \F_0}{\p t^{\al,m}}\frac{\p}{\p t^{\beta,m'}}+\frac{\p \F_0}{\p t^{\beta,m'}}\frac{\p}{\p t^{\al,m}}\biggr)
+\sum_{m,m'} b^{\al,m}_{\beta,m'} \tilde t^{\beta,m'}\frac{\p}{\p t^{\al,m}}.
\eeq
Taking the $x,t^{\gamma,0}$-derivatives on both sides of~\eqref{genus0-F0} we obtain
\begin{align}
&D(v_\gamma)+\sum_{m,m'} a^{\al,m;\beta,m'}
\bigl(\Omega^{[0]}_{1,0;\al,m}\Omega^{[0]}_{\gamma,0;\beta,m'}+ \Omega^{[0]}_{1,0;\beta,m'}\Omega^{[0]}_{\gamma,0;\al,m}\bigr)\nn\\
&\quad\quad\quad\quad + c_{1,0;\gamma,0}+ c_{\gamma,0;1,0} 
+\sum_{m} \bigl(b^{\al,m}_{1,0}\Omega^{[0]}_{\gamma,0;\al,m} + b^{\al,m}_{\gamma,0}\Omega^{[0]}_{1,0;\al,m}\bigr) = 0   \nn
\end{align}
and for $k\ge1$,
\begin{align}
&D(v_{\gamma,k})=\,\p_x D(v_{\gamma,k-1} )-\sum_{m,m'} a^{\al,m;\beta,m'} \Bigl(\Omega^{[0]}_{1,0;\al,m}\frac{\p v_{\gamma,k-1}}{\p t^{\beta,m'}}+\Omega^{[0]}_{1,0;\beta,m'}\frac{\p v_{\gamma,k-1}}{\p t^{\al,m}}\Bigr)\nn\\
&\quad\quad\quad\quad\quad +\sum_{m} b^{\al,m}_{1,0} \frac{\p v_{\gamma,k-1}}{\p t^{\al,m}}.\nn
\end{align}
It follows that 
for $k\geq1$,  
\beq\label{deg-D-vk}
\overline{\deg}\, D (v_{\gamma,k}) \leq k-1.
\eeq

Equations~\eqref{HHgs} can be equivalently written as follows:
\begin{align}
\frac{\p \HH_g}{\p s}=\,&D( \HH_g)+\sum_{m,m'} a^{\al,m;\beta,m'} \biggl(\,\sum_{k=1}^{g-1} \frac{\p \HH_k}{\p t^{\al,m}}\frac{\p\HH_{g-k}}{\p t^{\beta,m'}}+\frac{\p^2 \HH_{g-1}}{\p t^{\al,m}\p t^{\beta,m'}}\biggr),\quad g\geq 1. \label{HHgsD}
\end{align}

Write 
$$\HH_g=\sum_{\ell\geq 0} \HH_1^{[\ell]} s^\ell,\quad \HH_g^{[0]}=\F_g,\quad g\ge0.$$ 
By comparing coefficients of powers of~$s$ on both sides of~\eqref{HHgsD}
we find that~\eqref{HHgsD} with $g=1$ is equivalent to
\begin{align}
\HH_1^{[1]}=\,&D (\F_1)+\sum_{m,m'} a^{\al,m;\beta,m'} \frac{\p^2 \F_0}{\p t^{\al,m}\p t^{\beta,m'}}\nn\\
=\,& \sum_{k=0}^1 \frac{\p F_1}{\p v_{\gamma,k}}D( v_{\gamma,k})+\sum_{m,m'} a^{\al,m;\beta,m'}\Omega^{[0]}_{\al,m;\beta,m'}, \label{genus1Hl1}\\
\ell\,\HH_1^{[\ell]}=\,&D\bigl( \HH_1^{[\ell-1]} \bigr) 
,\quad \ell\geq 2. \label{genus1Hl2}
\end{align}
Similarly, equation~\eqref{HHgsD}  with $g\ge2$ is equivalent to
\begin{align}
\ell\,\HH_g^{[\ell]}=\,&D (\HH_g^{[\ell-1]}) 
+\sum_{m,m'} a^{\al,m;\beta,m'} \biggl(\,\sum_{k=1}^{g-1} \sum_{r=0}^{\ell-1} \frac{\p \HH_k^{[r]}}{\p t^{\al,m}}\frac{\p\HH_{g-k}^{[\ell-1-r]}}{\p t^{\beta,m'}}+\frac{\p^2 \HH_{g-1}^{[\ell-1]}}{\p t^{\al,m}\p t^{\beta,m'}}\biggr),\quad \ell\geq 1. \label{genusgHl}
\end{align}
The statements follow from~\eqref{deg-D-vk} and \eqref{genus1Hl1}--\eqref{genusgHl}.
\end{prfn}

\section{Integrable deformations having a tau-structure} \label{section4}
In this section we study properties for integrable deformations having a tau-structure, and prove Theorem~\ref{ab}.

Following the terminology of Section~\ref{Sect1}, let~\eqref{pert-int-intro} be an integrable deformation of the principal 
hierarchy of divergence form having a tau-structure. 
 Taking $(\beta,m)=(\alpha,0)$ in~\eqref{propertya} we obtain
 \beq\label{propertyaprime}
 \frac{\p \Omega_{1,0;\alpha,0}}{\p t^{\gamma,m'}} = \frac{\p\widehat\Omega_{\alpha,0;\gamma,m'}}{\p x}.
 \eeq
  Together with~\eqref{propertyb} and~\eqref{pert-int-intro} we get 
\beq\label{dOO}
\frac{\p\Omega_{\alpha,0;\gamma,m'}}{\p x}= \frac{\p\widehat\Omega_{\alpha,0;\gamma,m'}}{\p x}.
\eeq
Using the flow commutativity~\eqref{flowscomm} with $\alpha=1$
and using~\eqref{propertya}, we find 
\beq\label{ddsymm}
\p_x^2 \bigl(\widehat \Omega_{\beta,m;\gamma,m'}\bigr)=\p_x^2 \bigl(\widehat \Omega_{\gamma,m';\beta,m}\bigr),\quad m,m'\ge0.
\eeq
From the flow commutativity~\eqref{flowscomm} we get 
$$\p_{t^{\gamma,m'}}\p_{t^{\beta,m}} \bigl(\Omega_{1,0;\alpha,m''}\bigr)
=\p_{t^{\beta,m'}}\p_{t^{\gamma,m'}} \bigl(\Omega_{1,0;\alpha,m''}\bigr), 
\quad m,m',m''\ge0,$$
which can imply
\beq\label{dtaustruderiv}
\frac{\p}{\p x} \biggl(\frac{\p \widehat\Omega_{\alpha,m'';\beta,m}}{ \p t^{\gamma,m'}} \biggr)
=\frac{\p}{\p x} \biggl(\frac{\p \widehat\Omega_{\alpha,m'';\gamma,m'}}{ \p t^{\beta,m}} \biggr).
\eeq
It follows from \eqref{dOO}, \eqref{ddsymm}, \eqref{dtaustruderiv} that 
we can choose $\widehat \Omega_{\beta,m;\gamma,m'}$ so that for all $m,m',m''\ge0$,
\begin{align}
&\widehat \Omega_{\beta,m;\gamma,m'} = \widehat \Omega_{\gamma,m';\beta,m},\label{taustruc1}\\
&\frac{\p \widehat\Omega_{\alpha,m'';\beta,m}}{\p t^{\gamma,m'}}=\frac{\p\widehat\Omega_{\alpha,m'';\gamma,m'}}{\p t^{\beta,m}} , \label{taustruc2}\\
&\widehat \Omega_{\alpha,0;\beta,m} =  \Omega_{\alpha,0;\beta,m}.\label{OO}
\end{align}
By using~\eqref{propertya} and~\eqref{taustr0-2} we know that   
$$
\frac{\p \widehat\Omega^{[0]}_{\beta,m;\gamma,m'}({\bf w})}{\p x}=\frac{\p \Omega^{[0]}_{\beta,m;\gamma,m'}({\bf w})}{\p x},\quad m,m'\ge0.
$$
Using again the relations~\eqref{taustr0-2} we can fix the choice of $\widehat \Omega_{\alpha,m'';\beta,m}$ by requiring that 
$\widehat \Omega^{[0]}_{\beta,m;\gamma,m'}({\bf w})=\Omega^{[0]}_{\beta,m;\gamma,m'}({\bf w})$.
Because of~\eqref{OO} we can and will write $\widehat \Omega_{\alpha,m'';\beta,m}$ 
simply as 
$\Omega_{\alpha,m'';\beta,m}$\footnote{We shall note that the condition~b) in Definition~\ref{def-tau-symm} could be dropped, 
but then one can easily perform a Miura-type transformation to get back to the situation with~b).}.
We call $\Omega_{\alpha,m'';\beta,m}$ the {\it tau-structure} (or the {\it two-point correlations functions}).

Equations~\eqref{taustruc1}, \eqref{taustruc2} imply that for any solution ${\bf w}(\bt;\e)$ to the hierarchy~\eqref{pert-int-intro} 
 there exists a function $\tau=\tau(\bt;\e)$ such that
$$\e^2\frac{\p^2 \log\tau(\bt;\e)}{\p t^{\alpha,m''} \p t^{\beta,m}}
=\Omega_{\alpha,m'';\beta,m}\Bigl({\bf w}(\bt;\e), \frac{\p {\bf w}(\bt;\e)}{\p x}, \cdots;\e\Bigr),\quad m,m''\ge0.$$
The function $\tau$ is called the {\em tau-function} of the solution~${\bf w}(\bt;\e)$.

\begin{rmk}
In Definition~\ref{def-tau-symm}, we have required that the dispersionless part of the integrable hierarchy 
to be coincide with the principal hierarchy of the Frobenius manifold. It is very natural to drop this requirement and extend the  
definition to the dispersionless part allowing non-hamiltonian dispersionless tau-structure, in particular on primary two-point correlation functions. 
However, further axioms are needed to guarantee integrability (which means here the reconstruction of commuting descendent flows).
It will be interesting if the approach from~\cite{BR24} for dispersionless KP hierarchy regarding two-point functions 
could be helpful for us, and it will also be interesting if dispersionless tau-structure could be connected to geometry studied in~\cite{Shima, Totaro}.
\end{rmk}

\begin{dfn}
A coordinate change ${\bf w}\rightarrow \widetilde {\bf w}$ of the form
\begin{equation}\label{normal-trans}
\widetilde{w}_\al=w_\al+\e^2 \p_x\p_{t^{\al,0}} \Bigl(\,\sum_{k\ge 0} \e^{2k} A_{k}\Bigr),
\end{equation}
together with the following change of the tau-structure 
\beq
\widetilde{\Omega}_{\alpha,m'';\beta,m}:=\Omega_{\alpha,m'';\beta,m}+\e^2\p_{t^{\alpha,m''}} \p_{t^{\beta,m}} \Bigl(\,\sum_{k\geq 0} \e^{2k} A_{k}\Bigr)
\eeq
is called a normal Miura-type transformation if
$A_{k}\in \mathcal{A}_{\bf w}^{[2k]}$. The coordinate change~\eqref{normal-trans} is called 
normal quasi-Miura if $A_k$ 
 depend rationally on $w^\gamma_1$ and 
 polynomially on $w^\gamma_\ell$ ($\ell\ge 2$) with $\deg A_k=2k$.
\end{dfn}

By a direct verification, 
a normal Miura-type transformation transforms an integrable deformation of the principal hierarchy
of divergence form having a tau-structure to another one of the same type. (This is similar to \cite[Lemma~4.3]{DLYZ}.)
Generally speaking, a normal quasi-Miura transformation does not preserve polynomiality. 

Recall that the two-point correlation functions
for the integrable hierarchy of topological type are defined by
$$
\Omega^{{\rm DZ, flat}}_{\beta,m; \gamma,m'}({\bf v}, {\bf v}_1, {\bf v}_2,\dots;\e)
=\Omega^{[0], {\rm flat}}_{\beta,m; \gamma,m'}({\bf v})
+\sum_{g\geq 1} \e^{2g} \frac{\p F_g({\bf v}, {\bf v}_1, \dots, {\bf v}_{3g-2})}{\p t^{\beta,m} \p t^{\gamma,m'}}, \quad m,m'\ge0.
$$
Substituting the inverse of~\eqref{quasitrivialmap} 
in $\Omega^{{\rm DZ, flat}}_{\beta,m; \gamma,m'}({\bf v}, {\bf v}_1, {\bf v}_2,\dots;\e)$ 
one obtains the two-point correlation functions in terms of the normal coordinates ${\bf u}$ and their jets $ {\bf u}_1, {\bf u}_2$, \dots,  
 which are denoted by $\Omega^{{\rm DZ}}_{\beta,m; \gamma,m'}({\bf u}, {\bf u}_1, {\bf u}_2,\dots;\e)$.

Let $L$ be an operator of the form~\eqref{def-L} satisfying~\eqref{zero-const-L}, and 
consider the $L$-deformed hierarchy (see Section~\ref{Sect1}). 
Since $\HH_g(\bt;s=0)=\F_g(\bt)$, it is obvious that 
the $L$-deformed hierarchy is also a deformation of the integrable hierarchy of 
topological type. 
Define the two-point correlation functions of the $L$-deformed hierarchy by
$$
\Omega^{L, {\rm flat}}_{\beta,m;\gamma,m'}({\bf v}, {\bf v}_1, {\bf v}_2,\dots;s,\e)
=\Omega^{[0], {\rm flat}}_{\beta,m; \gamma,m'}({\bf v})
+\sum_{g\geq 1} \e^{2g} \frac{\p H_g({\bf v}, {\bf v}_1, \dots, {\bf v}_{3g-2};s)}{\p t^{\beta,m} \p t^{\gamma,m'}}.$$
Then we have
$$\Omega^{L, {\rm flat}}_{\beta,m;\gamma,m'}({\bf v}, {\bf v}_1, {\bf v}_2,\dots;s,\e)
=\Omega^{{\rm DZ, flat}}_{\beta,m; \gamma,m'}({\bf v}, {\bf v}_1, {\bf v}_2,\dots;\e)
+\e^2\frac{\p^2 (\HH-\F)}{\p t^{\beta,m} \p t^{\gamma,m'}},$$
where we recall that $\HH$ and $\F$ are given in~\eqref{der-e-H}--\eqref{genus-expan-H} and~\eqref{genusF}, respectively.
Substituting the inverse of the following quasi-trivial map 
$$w_\alpha=v_\alpha+\e^2 +\sum_{g\geq 1} \e^{2g} \frac{\p H_g({\bf v}, {\bf v}_1, \dots, {\bf v}_{3g-2};s)}{\p x \p t^{\alpha,0}}$$
into 
$\Omega^{L, {\rm flat}}_{\beta,m;\gamma,m'}({\bf v}, {\bf v}_1, {\bf v}_2,\dots;s,\e)$, 
one obtains the two-point correlation functions in the normal coordinates ${\bf w}$ and their jets ${\bf w}_1, {\bf w}_2$, \dots, which 
are denoted by $$\Omega^L_{\beta,m;\gamma,m'}({\bf w}, {\bf w}_1, {\bf w}_2,\dots;s,\e).$$
The $L$-deformed hierarchy, by definition, can be written as
\beq\label{Ldeformedhier}
\frac{\p w_\alpha}{\p t^{\beta,m}}=\frac{\p \Omega^L_{\alpha,0;\beta,m}({\bf w}, {\bf w}_1, {\bf w}_2,\dots;s,\e)}{\p x},\quad m\geq 0.
\eeq

We are ready to prove Theorem~\ref{ab}.

\begin{prfn}{Theorem~\ref{ab}} 
By construction the flows of the $L$-deformed hierarchy~\eqref{Ldeformedhier} (which are 
of divergence form) commute pairwise, and are 
deformations of the principal hierarchy~\eqref{ph-intro}.
Moreover, it is easy to see that $\Omega^L_{\alpha,0;1,0}=\Omega^L_{1,0;\alpha,0}=w_\alpha$ and  
$$
\frac{\p \Omega^L_{1,0;\beta,m}}{\p t^{\gamma,m'}}=\frac{\p \Omega^L_{\beta,m;\gamma,m'}}{\p x},\quad m,m'\ge0.
$$

It remains to prove that $\Omega^L_{\gamma,p;\mu,q}({\bf w}, {\bf w}_1, {\bf w}_2,\dots;s,\e)$ have polynomiality.

First of all, the chain rule tells that for all $p,q\ge0$,
\begin{align}
&\quad\quad\frac{\p \Omega^L_{\gamma,p;\mu,q}({\bf w}, {\bf w}_1, {\bf w}_2,\dots;s,\e)}{\p s}\nn\\
&=\,\frac{\p \Omega^{L,{\rm flat}}_{\gamma,p;\mu,q}({\bf v}, {\bf v}_1, {\bf v}_2,\dots;s,\e)}{\p s}
-\sum_{k\geq 0} \frac{\p \Omega^L_{\gamma,p;\mu,q}({\bf w}, {\bf w}_1, {\bf w}_2,\dots;s,\e)}{\p w_{\phi,k}}\p_x^k\Bigl(\frac{\p w_{\phi}}{\p s}\Bigr)\nn\\
&=\,\e^2\frac{\p^2}{\p t^{\gamma,p}\p t^{\mu,q}}\Bigl(\frac{\p \HH}{\p s}\Bigr)
-\sum_{k\geq 0} \frac{\p \Omega^L_{\gamma,p;\mu,q}({\bf w}, {\bf w}_1, {\bf w}_2,\dots;s,\e)}{\p w_{\phi,k}} \p_x^k \Bigl(\frac{\p w_{\phi}}{\p s}\Bigr).
\label{chain-rule}
\end{align}

Let us simplify the right-hand side of~\eqref{chain-rule}. Recall that $\HH$ satisfies
\begin{align}
\frac{\p \HH}{\p s}=&~\e^2\sum_{m, m'} a^{\al,m;\beta,m'} \left(\frac{\p \HH}{\p t^{\al,m}}\frac{\p\HH}{\p t^{\beta,n}}+\frac{\p^2 \HH}{\p t^{\al,m}\p t^{\beta,m'}}\right)+\sum_{m,m'} b^{\al,m}_{\beta,m'} \tilde t^{\beta,m'}\frac{\p \HH }{\p t^{\al,m}}\nn\\
&+\frac{1}{\e^2}\sum_{m,m'} c_{\al,m;\beta,m'}\tilde t^{\al,m}\tilde t^{\beta,m'}.\nn
\end{align}
Therefore, 
\begin{align}
&\frac{\p^2}{\p t^{\gamma,p}\p t^{\mu,q}}\Bigl(\frac{\p \HH}{\p s}\Bigr)\nn\\
&=\sum_{m,m'} a^{\al,m;\beta,m'}\left(\frac{\p^2 \Omega^{L, {\rm flat}}_{\al,m;\beta,m'}}{\p t^{\gamma,p}\p t^{\mu,q}}
+\frac{1}{\e^2}\Omega^{L, {\rm flat}}_{\al,m;\gamma,p} \Omega^{L, {\rm flat}}_{\beta,m';\mu,q}
+\frac{1}{\e^2}\Omega^{L, {\rm flat}}_{\al,m;\mu,q} \Omega^{L, {\rm flat}}_{\beta,m';\gamma,p}\right)\nn\\
&+\sum_{m,m'} a^{\al,m;\beta,m'} \left(\frac{\p \Omega^{L, {\rm flat}}_{\gamma,p;\mu,q}}{\p t^{\al,m}}\frac{\p\HH}{\p t^{\beta,m'}}
+\frac{\p \Omega^{L, {\rm flat}}_{\gamma,p;\mu,q}}{\p t^{\beta,m'}}\frac{\p\HH}{\p t^{\al,m}}\right)\nn\\
&+\frac{1}{\e^2}\sum_{m} \left(b^{\al,m}_{\gamma,p} \Omega^{L, {\rm flat}}_{\al,m;\mu,q}+b^{\al,m}_{\mu,q}\Omega^{L, {\rm flat}}_{\al,m;\gamma,p}\right)
+\frac{1}{\e^2}\sum_{m,m'} b^{\al,m}_{\beta,m'} \tilde t^{\beta,m'}\frac{\p \Omega^{L, {\rm flat}}_{\gamma,p;\mu,q} }{\p t^{\al,m}}
\nn\\
&+\frac{1}{\e^2} \left(c_{\gamma,p;\mu,q}+c_{\mu,q;\gamma,p}\right).\nn
\end{align}
Substituting this expression in equation \eqref{chain-rule} we find
\begin{align}
&\frac{\p \Omega^L_{\gamma,p;\mu,q}}{\p s}\nn\\
&=\sum_{m,m'} a^{\al,m;\beta,m'}\left(\e^2\frac{\p^2 \Omega^{L, {\rm flat}}_{\al,m;\beta,m'}}{\p t^{\gamma,p}\p t^{\mu,q}}
+\Omega^{L, {\rm flat}}_{\al,m;\gamma,p} \Omega^{L, {\rm flat}}_{\beta,m';\mu,q}+\Omega^{L, {\rm flat}}_{\al,m;\mu,q} \Omega^{L, {\rm flat}}_{\beta,m';\gamma,p}\right)\nn\\
&+\,\e^2\sum_{m,m'} a^{\al,m;\beta,m'} \left(\frac{\p \Omega^{L, {\rm flat}}_{\gamma,p;\mu,q}}{\p t^{\al,m}}\frac{\p\HH}{\p t^{\beta,m'}}+\frac{\p \Omega^{L, {\rm flat}}_{\gamma,p;\mu,q}}{\p t^{\beta,m'}}\frac{\p\HH}{\p t^{\al,m}}\right)\nn\\
&+\sum_m\left(b^{\al,m}_{\gamma,p} \Omega^{L, {\rm flat}}_{\al,m;\mu,q}+b^{\al,m}_{\mu,q}\Omega^{L, {\rm flat}}_{\al,m;\gamma,p}\right)+\sum_{m,m'} b^{\al,m}_{\beta,m'} \tilde t^{\beta,m'}\frac{\p \Omega^{L, {\rm flat}}_{\gamma,p;\mu,q} }{\p t^{\al,m}}\nn\\
&+c_{\gamma,p;\mu,q}+c_{\mu,q;\gamma,p}-\sum_{k\geq 0} \frac{\p \Omega^L_{\gamma,p;\mu,q}}{\p w_{\phi,k}}\p_x^k\Bigl(\frac{\p w_{\phi}}{\p s}\Bigr).\nn
\end{align}
In particular, taking in the above formula $\gamma=\phi,\mu=1,p=q=0$ and noticing that $\Omega^L_{\phi,1;1,0}=w_\phi,$ we have
\begin{align}
0=&\,\sum_{m,m'} a^{\al,m;\beta,m'}\left(\e^2\frac{\p^2 w_\phi}{\p t^{\al,m}\p t^{\beta,m'}}
+\Omega^{L, {\rm flat}}_{\al,m;\phi,0} \Omega^{L, {\rm flat}}_{\beta,m';1,0}
+\Omega^{L, {\rm flat}}_{\al,m;1,0} \Omega^{L, {\rm flat}}_{\beta,m';\phi,0}\right)\nn\\
&+\e^2\sum_{m,m'} a^{\al,m;\beta,m'} \biggl(\frac{\p w_\phi}{\p t^{\al,m}}\frac{\p\HH}{\p t^{\beta,m'}}
+\frac{\p \Omega^{L, {\rm flat}}_{\phi,0;1,0}}{\p t^{\beta,m'}}\frac{\p\HH}{\p t^{\al,m}}\biggr)\nn\\
&+\sum_m \left(b^{\al,m}_{\phi,0} \Omega^{L, {\rm flat}}_{\al,m;1,0}+b^{\al,m}_{1,0}\Omega^{L, {\rm flat}}_{\al,m;\phi,0}\right)+\sum_{m,m'} b^{\al,m}_{\beta,m'} \tilde t^{\beta,n}\frac{\p w_\phi }{\p t^{\al,m}}\nn\\
&+c_{\phi,0;1,0}+ c_{1,0;\phi,0}-\frac{\p w_{\phi}}{\p s}.\nn
\end{align}

It follows that 
\begin{align}
&\frac{\p \Omega^L_{\gamma,p;\mu,q}}{\p s}\nn\\
&=\sum_{m,m'} a^{\al,m;\beta,m'}\left(\e^2\frac{\p^2 \Omega^{L, {\rm flat}}_{\al,m;\beta,m'}}{\p t^{\gamma,p}\p t^{\mu,q}}
+\Omega^{L, {\rm flat}}_{\al,m;\gamma,p} \Omega^{L, {\rm flat}}_{\beta,m';\mu,q}
+\Omega^{L, {\rm flat}}_{\al,m;\mu,q} \Omega^{L, {\rm flat}}_{\beta,m';\gamma,p}\right)\nn\\
&+\,\e^2\sum_{m,m'} a^{\al,m;\beta,m'} \left(\frac{\p \Omega^{L, {\rm flat}}_{\gamma,p;\mu,q}}{\p t^{\al,m}}\frac{\p\HH}{\p t^{\beta,m'}}
+\frac{\p \Omega^{L, {\rm flat}}_{\gamma,p;\mu,q}}{\p t^{\beta,m'}}\frac{\p\HH}{\p t^{\al,m}}\right)\nn\\
&+\sum_m\left(b^{\al,m}_{\gamma,p} \Omega^{L, {\rm flat}}_{\al,m;\mu,q}+b^{\al,m}_{\mu,q}\Omega^{L, {\rm flat}}_{\al,m;\gamma,p}\right)+\sum_{m,m'} b^{\al,m}_{\beta,m'} \tilde t^{\beta,m'}\frac{\p \Omega^{L, {\rm flat}}_{\gamma,p;\mu,q} }{\p t^{\al,m}}
+c_{\gamma,p;\mu,q}+c_{\mu,q;\gamma,p}\nn\\
&-\sum_{k\geq 0} \frac{\p \Omega^L_{\gamma,p;\mu,q}}{\p w_{\phi,k}}\p_x^k \left(\,\sum_{m,m'} a^{\al,m;\beta,m'}\left(\e^2\frac{\p^2 w_\phi}{\p t^{\al,m}\p t^{\beta,m'}}+\Omega^{L, {\rm flat}}_{\al,m;\phi,0} \Omega^{L, {\rm flat}}_{\beta,m';1,0}+\Omega^{L, {\rm flat}}_{\al,m;1,0} \Omega^{L, {\rm flat}}_{\beta,m';\phi,0}\right)\right.\nn\\
&\left.+\,\e^2\sum_{m,m'} a^{\al,m;\beta,m'} \left(\frac{\p w_\phi}{\p t^{\al,m}}\frac{\p\HH}{\p t^{\beta,m'}}+\frac{\p w_\phi}{\p t^{\beta,m'}}\frac{\p\HH}{\p t^{\al,m}}\right)\right.\nn\\
&\left.+\sum_m\left(b^{\al,m}_{\phi,0} \Omega^{L, {\rm flat}}_{\al,m;1,0}+b^{\al,m}_{1,0}\Omega^{L, {\rm flat}}_{\al,m;\phi,0}\right)
+\sum_{m,m'} b^{\al,m}_{\beta,m'} \tilde t^{\beta,m'}\frac{\p w_\phi }{\p t^{\al,m}} +c_{\phi,0;1,0}+ c_{1,0;\phi,0}
\right).\nn
\end{align}
Simplifying the above formula we arrive at 
\begin{align}
&\frac{\p \Omega^L_{\gamma,p;\mu,q}}{\p s}\nn\\
&=\sum_{m,m'} a^{\al,m;\beta,m'}\biggl(\e^2\frac{\p^2 \Omega^L_{\al,m;\beta,m'}}{\p t^{\gamma,p}\p t^{\mu,q}}
+\Omega^L_{\al,m;\gamma,p} \Omega^L_{\beta,m';\mu,q}+\Omega^L_{\al,m;\mu,q} \Omega^L_{\beta,m';\gamma,p}\biggr)\nn\\
&+\sum_{m} \left(b^{\al,m}_{\gamma,p} \Omega^L_{\al,m;\mu,q}+b^{\al,m}_{\mu,q}\Omega^L_{\al,m;\gamma,p}\right)+c_{\gamma,p;\mu,q}+c_{\mu,q;\gamma,p}\nn\\
&-\sum_{k\geq 0} \frac{\p \Omega^L_{\gamma,p;\mu,q}}{\p w_{\phi,k}}\p_x^k \left(\sum_{m,m'} a^{\al,m;\beta,m'}\left(\e^2\frac{\p^2 w_\phi}{\p t^{\al,m}\p t^{\beta,m'}}+\Omega^L_{\al,m;\phi,0} \Omega^L_{\beta,m';1,0}
+\Omega^L_{\al,m;1,0} \Omega^L_{\beta,m';\phi,0}\right)\right.\nn\\
&\left.+\sum_{m} \left(b^{\al,m}_{\phi,0} \Omega^L_{\al,m;1,0}+b^{\al,m}_{1,0}\Omega^L_{\al,m;\phi,0}\right)+c_{\phi,0;1,0}+ c_{1,0;\phi,0}\right)\nn\\
& - \e^2 \sum_{k\geq 0} \frac{\p \Omega^L_{\gamma,p;\mu,q}}{\p w_{\phi,k}}\sum_{m,m'} a^{\al,m;\beta,m'} 
\sum_{j=1}^{k} \binom{k}{j} \left( \p_x^{k-j} \Bigl(\frac{\p w_\phi}{\p t^{\al,m}}\Bigr) \p_x^{j-1} (\Omega^L_{\beta,m';1,0}) \right.\nn\\
& \qquad \qquad \left. +\p_x^{k-j}\Bigl(\frac{\p w_\phi}{\p t^{\beta,m'}}\Bigr) \p_x^{j-1}(\Omega^L_{\al,m;1,0})\right)  \nn \\
& - \sum_{m\geq 0, k\ge1} k b^{\alpha,m}_{1,0} \frac{\p \Omega^L_{\gamma,p;\mu,q}}{\p w_{\phi,k}} \p_x^{k-1} \Bigl(\frac{\p w_\phi }{\p t^{\al,m}}\Bigr).\nn
\end{align}
The theorem is then proved by noticing that the initial values 
$$
 \Omega^L_{\gamma;p;\mu,q}({\bf w}, {\bf w}_1, {\bf w}_2,\dots;s=0,\e)
=\Omega^{\rm DZ}_{\gamma;p;\mu,q}({\bf w}, {\bf w}_1, {\bf w}_2,\dots;\e)
$$
have~\cite{BPS1,BPS2}  polynomiality.
\end{prfn}

\section{The Virasoro-like hierarchy} \label{section5}
In this section, employing Theorem~B and Theorem~\ref{ab} we give more details about the construction of the Virasoro-like hierarchy.

It was found in~\cite{LYZZ} that $L_i(\nu)$, $i\ge-1$, are even polynomials of~$\nu$ and we can write 
\begin{equation}
L_i(\nu)=:\sum_{j=0}^{[\f{i+1}{2}]} \,L_{i,2j}\nu^{2j}, \quad i\ge-1.
\end{equation}
The operators $L_{i,2j}$, $1\leq j \leq \left[\frac{i+1}2\right]$, are the {\it Virasoro-like operators}, which have been briefly 
discussed in Section~\ref{Sect1}.
Obviously, for $i\geq -1$ we have $L_{i,0}= L_i$.

\begin{rmk}
In view of the Givental quantization~\cite{G1}, 
\beq\label{G1-1}
L_i(\nu)= \widehat{l_i(\nu)} + \frac{\delta_{i,0}}{4} {\rm tr} \Bigl(\frac14 - \mu^2\Bigr), \quad i\ge0,
\eeq
where 
\beq\label{G1-2}
l_i(\nu) : = \frac12 z^{-\frac12} \Bigl((z \circ \partial_z \circ z - (\mu+\nu) z + R)^{i+1} + (z \circ \partial_z \circ z - (\mu-\nu) z + R)^{i+1}\Bigr) z^{-\frac12}
\eeq
with $\mu,R$ being the two constant matrices given in Section~\ref{section2}. We see that the polynomiality in~$\nu$ 
is more obvious from~\eqref{G1-1} and~\eqref{G1-2}.
We note that the expressions of $l_i(0)$ and of the top coefficients of~$l_{2j-1}(\nu)$ are 
discussed in~\cite{G1}, which are symplectic. But in general coefficients of $l_i(\nu)$ give rise to non-symplectic infinitesimal actions.
\end{rmk}

By using Theorem~B and Theorem~\ref{ab}, one can obtain  
integrable deformations of the principal hierarchy of divergence form having a tau-structure. 

For any $i\geq -1$, the Virasoro operator $L_i=L_{i,0}$ is an operator of the form \eqref{def-L} satisfying \eqref{zero-const-L}, 
and the $L_i$-deformed hierarchy gives a trivial deformation as 
it just coincides with the integrable hierarchy of topological type.
For non-trivial deformations, we shall use the Virasoro-like operators $L_{i,2j}$ with $j\ge1$.
To get the most general deformation from these operators, a direct way is to consider the following operator 
\beq\label{defLuni}
L_{\rm universal} = \sum_{i\ge1} \sum_{j=1}^{[(i+1)/2]} a_{i,2j} L_{i,2j},
\eeq
with $a_{i,2j}$ being indeterminates. 
Define
\beq
\widehat{Z}_{\rm like} = \widehat{Z}_{\rm like} (\bt; {\bf a};\e) := e^{L_{\rm universal}} (Z),
\eeq
where ${\bf a}=(a_{1,2}, a_{2,2}, a_{3,2},\dots)$ is an infinite sequence. 
The Dubrovin--Zhang hierarchy associated to $\widehat Z_{\rm like}$ will be called the 
 {\it Virasoro-like hierarchy obtained from $L_{\rm universal}$}.
 \begin{cor}
The {\it Virasoro-like hierarchy obtained from $L_{\rm universal}$}
is an integrable deformation of the principal 
hierarchy of divergence form having a tau-structure.
\end{cor}
\begin{prf}
By using Theorem~\ref{ab} with $L=L_{\rm universal}$ and putting at the end $s=1$.
\end{prf}

For a practical reason, we can also successively apply the Virasoro-like operators 
 in a certain ordering on the partition function as it was done in Section~\ref{Sect1}.

\begin{prfn}{Theorem~\ref{B-o}}
When we successively apply Virasoro-like operators in a certain order on $Z$, 
we get a sequence of partition functions and a sequence of deformations of the principal hierarchy. 
By induction (using the proofs similar to those of Theorem~\ref{H} and Theorem~\ref{ab}) 
we can show that these partition functions all admit genus expansions 
whose genus~0 parts equal $\F_0(\bt)$ and whose genus $g$ ($g\ge1$) 
parts depend on the jets up to ${\bf v}_{3g-2}$, and these deformations are all 
 integrable deformations of the principal 
hierarchy of divergence form having a tau-structure.
\end{prfn}

\begin{rmk}
If we put 
\begin{align}
& a_{i,2j}=0,\quad \forall\, 1\leq j<\left[\frac{i+1}2\right],\nn\\
& a_{2k-1,2k}=s_k, \quad k\geq 1 \nn
\end{align}
in the Virasoro-like hierarchy obtained from $L_{\rm universal}$, 
or if we put 
\begin{align}
&  r_{i,2j}=0,\quad \forall\, 1\leq j<\left[\frac{i+1}2\right],\nn\\
&  r_{2k-1,2k}=s_k, \quad k\geq 1 \nn
\end{align}
in the Virasoro-like hierarchy via successive applications of Virasoro-like operators, 
we get the Hodge integrable hierarchy~\cite{DLYZ}. Here we recall that  
the operators $L_{2j-1,2j}$, $j\ge1$, all commute as we mentioned in Section~\ref{Sect1}, 
whereas in general the operators $L_{i,2j}$ do not commute. 
\end{rmk}

In the rest of this section, let us consider the case of the one-dimensional Frobenius manifold.
In this case, the principal hierarchy is the Riemann--Hopf hierarchy
\begin{equation}\label{Riemann}
\frac{\p v}{\p t_m}=\p_x \bigl(\Omega_{0;m}^{[0]}(v)\bigr),\quad m\geq 0,
\end{equation}
where 
\begin{align}\label{two-point-genus-zero-kdv}
\Omega_{0;m}^{[0]}(v)=\frac{v^{m+1}}{(m+1)!}.
\end{align}
Here we have made change of notations: $t^{1,m}\rightarrow t_m$, $\Omega_{1,0;1,m}\rightarrow \Omega_{0;m}$.

Let us first consider a particular example. 
\begin{emp}
{\bf The $L_{2,2}$-deformed hierarchy.} \quad The $L_{2,2}$ operator reads
$$L_{2,2}=\sum_{p\geq 0} \frac{3}{2}(2p+3)\,\tilde{t}_p \frac{\p }{\p t_{p+2}}-\frac{3}{2}\e^2\frac{\p^2}{\p t_0 \p t_1}.$$
Using the algorithm described in the proof of Theorem~\ref{H}, we obtain
\begin{align}
H_1&=\,\frac{1}{24} \log(v_1)-\frac{3}{4} s\, v^2,\nn\\
H_2&=\,\frac{v_{2}^3}{360 v_1^4}-\frac{7 v_{2} v_3}{1920 v_1^3}+\frac{v_4}{1152 v_1^2}+s \left(-\frac{5}{32} v_{2}+\frac{11 v v_{2}^2}{160 v_1^2}-\frac{3 v v_3}{40 v_1}\right)\nn\\
&\quad+s^2 \left(\frac{45}{16} v v_1^2+\frac{63}{40} v^2 v_{2}\right)+s^3\left(-\frac{27}{10} v^3 v_1^2\right),\nn
\end{align}
etc.
Substituting the normal quasi-Miura transformation
\begin{equation}\label{quasi-L22}
w=v+\sum_{g\geq 1}\e^{2g} \p_x^2(\HH_g)
\end{equation}
into the Riemann--Hopf hierarchy we obtain the $L_{2,2}$-deformed hierarchy:
\begin{align}
w_{t_0}&=\p_x(w),\nn\\
w_{t_1}&=\p_x\left[\frac{w^2}{2}+\e^2\Bigl(\,\frac{w_{2}}{12}-\frac{3}{2} s\, w\, w_1^2\Bigr)+\e^4 \left( s\, \Bigl(-\frac{1}{10} w_{2}^2-\frac{3}{10} w_1 w_3-\frac{1}{20} w w_4\Bigr)\right.\right.\nn\\
&\left.\left.\quad\quad +s^2 \Bigl(\,\frac{81}{16} w_1^4+\frac{171}{10} w w_1^2 w_{2}+\frac{18}{5} w^2 w_{2}^2+\frac{9}{5} w^2 w_1 w_3\Bigr)\right.\right.\nn\\
&\left.\left.\quad\quad +s^3 \Bigl(-\frac{243}{10} w^2 w_1^4-\frac{108}{5} w^3 w_1^2 w_{2}\Bigr)\right)+\e^6\Bigl(-\frac{1}{280} s \,w_6 +\cdots\Bigr)+\cdots\right],\nn
\end{align}
etc. Performing a hamiltonian test (cf.~\cite{DLYZ, DZ-norm, LWZ, LZ}), we can show that 
the $L_{2,2}$-deformed hierarchy 
does not have a polynomial hamiltonian structure. 
\end{emp}

We now consider, following Definition~\ref{def-tau-symm}, classification of 
integrable deformations of the Riemann--Hopf hierarchy of divergence form (cf.~\cite{ALM}) having a tau-structure.
Such an integrable deformation has the form
\begin{equation}
\label{general} \frac{\p w}{\p t_m}=\p_x (\Omega_{0;m}), \quad m\geq 0,
\end{equation}
where 
\begin{eqnarray}
\label{omop-polynomial} \Omega_{0;m}=\frac{w^{m+1}}{(m+1)!}+\sum_{g\geq 1}\e^{2g} \Omega_{0;m}^{[g]}, \quad 
\Omega_{0;m}^{[g]}\in \mathcal{A}_w^{[2g]}~(g\ge1),
\end{eqnarray}
and must satisfy other conditions in Definition~\ref{def-tau-symm}.

\begin{rmk}
Several special cases for one-component integrable systems with tau-structure have been 
considered in~\cite{DLYZ, DVY, LWZZ, YZ}. In particular, it was shown in~\cite{DVY, LWZZ} that the Sawada--Kotera hierarchy and 
the Kaup--Kupershmidt hierarchy for subfamilies of flows satisfy the above-mentioned axioms.
\end{rmk}

\begin{cnj}\label{standard-form}
Any integrable deformation of the Riemann--Hopf hierarchy of divergence form having a tau-structure is equivalent, modulo a normal Miura transformation, to the hierarchy which is uniquely specified by $\Omega_{0;1}$:
\begin{align}
&\Omega_{0;1}=\frac{w^2}{2}+\epsilon^2 a_0 w_2+\epsilon^4(a_1(w) w_4+ b_1(w) w_3w_1+b_2(w) w_1^4)\nn\\
&+\epsilon^6(a_2(w) w_1 w_5 + b_3(w) w_6 + b_4(w) w_4 w_1^2 + b_5(u) w_3w_1^3 + b_6(w) w_3^2 + b_7(w) w_1^6)
\nn\\
&+\e^8(a_3(w) w_1^2 w_6 + b_8(w) w_8 + b_9(w) w_1 w_7+b_{10}(w) w_3 w_5+ b_{11}(w) w_1^3 w_5+b_{12}(w) w_4^2 \nn\\
&+b_{13}(w)w_1 w_3 w_4 + b_{14}(w) w_1^4 w_4 + b_{15}(w)w_1^2w_3^2+b_{16}(w)w_1^5 w_3 + b_{17}(w) w_1^8)+\cdots.\label{stand-om01}
\end{align}
Here, $a_0$ must be a constant parameter, $a_1(w),b_1(w),a_2(w),b_2(w),\dots$ are smooth functional parameters, 
the coefficients of $\e^{2g}$ with $g\ge2$ do not contain $w_2$, and $a_i(w)$ are coefficients of $$\e^{2i+2}w_1^{i-1} w_{i+3}, \quad i\ge1.$$ 
Moreover, $b_1(w),b_2(w),\dots$ are uniquely determined by $a_1(w),a_2(w),\dots$.
\end{cnj}
Note that by a rescaling of $\e$ we can and will assume $$a_0=\frac{1}{12}.$$
The first few expressions of $b_i$ are then given by
\begin{align}
&b_1=\frac{5}{6}a_1',\nn\\
&b_2=\frac{1}{16}a_1''',\nn\\
&b_3=\frac{10}{7}a_1^2+\frac{1}{14} \,a_1'. \nn
\end{align}

For the $a_{1,2}L_{1,2}+a_{2,2}L_{2,2}+a_{3,4} L_{3,4}$-deformed hierarchy we have
\begin{align}
&a_1=\frac{ a_{1,2} }{60}-\frac{a_{2,2}}{20} u,\nn\\
&a_2=\frac{a_{3,4} }{720} +\frac{a_{1,2}^3}{150} +\frac{7a_{1,2}a_{2,2}}{400} + \Bigl(\,\frac{3a_{1,2}^2a_{2,2}}{50} +\frac{3a_{2,2}^2 }{400} \Bigr)u+\frac{9a_{1,2}a_{2,2}^2}{50} u^2+\frac{9a_{2,2}^3}{50} u^3.\nn
\end{align}

\begin{cnj} \label{un-cnj} 
The Virasoro-like hierarchy of the one-dimensional Frobenius manifold is a universal object in 
integrable deformations of the Riemann--Hopf hierarchy having a tau-structure.
\end{cnj}

We also have a stronger version of the above Conjecture~\ref{un-cnj}: the constant parameter $a_{i,2j}$ (see~\eqref{defLuni}) provides a free
 parameter for the coefficient of $u^{i-2j+1}$ in the Taylor expansion of~$a_j(u)$
and in such a representation $a_j(u)$ is homogeneous of degree $2j-1$ with respect to the degree
assignments: $\deg a_{i,2j}=i$ and $\deg u=-1$.

We end the paper with a brief discussion about the Sawada--Kotera hierarchy and how it fits into the standard form 
in Conjecture~\ref{standard-form}.
The first few flows of the Sawada--Kotera hierarchy are given by
\begin{align} 
& \frac{\p u}{\p T_0} = \p_x (u) ,\nn\\
& \frac{\p u}{\p T_1} = \p_x \Bigl(\, \frac{u^2}{2}  +\e^2 \Bigl( \frac{u''}{2}-\frac{u'^2}{4 u} \Bigr)+  \e^4 \Bigl( -\frac{u^{(4)}}{20 u}
+\frac{u''^2}{5 u^2}+\frac{17 u'^4}{40 u^4}+\frac{7 u^{(3)} u'}{20 u^2}-\frac{19 u'^2 u''}{20 u^3} \Bigr) + \mathcal{O}(\e^6) \Bigr) , \label{firstflow} \\
& \frac{\p u}{\p T_2} = \p_x \Bigl(\frac{u^3}{6}  +\frac{\e^2}{2}  u \, u''+ \frac{\e^4}{10}  u^{(4)}\Bigr) \nn \qquad ({\rm Sawada-Kotera~equation})
\end{align}
(the normalization is slightly different from usual), where prime means derivative with respect to~$x$. Under the following normal Miura transformation 
\beq
w =  u -1 +\p_x^2  \Bigl(  \frac{\e^2}{4} \log u + \e^4 \Bigl( \, \frac{5 u'^2}{48 u^3}-\frac{67 u''}{480 u^2}\Bigr) + \mathcal{O}(\e^6)  \Bigr)  ,
\eeq
equation~\eqref{firstflow} becomes
\beq
w_{T_1} = \p_x \Bigl( \frac{2w+w^2}2  + \frac{\e^2}{2}  w'' 
+ \e^4 \Bigl(\frac{3 w'^4}{160 (1+w)^4}+\frac{w^{(3)} w'}{24 (1+w)^2}-\frac{w^{(4)}}{20 (1+w)}\Bigr) + \mathcal{O}(\e^6) \Bigr) \,.
\eeq
To match with the normal form in Conjecture~\ref{standard-form}, we consider $\p_{T_1}=\p_{t_1}+ \p_{t_0}$ and put $\e^2\mapsto \e^2/6$, and 
get
\beq
a_{1,2}= -\frac1{12}, \quad a_{2,2}= -\frac1{36}.
\eeq


\begin{thebibliography}{99}

\bibitem{ALM}
Arsie, A., Lorenzoni, P., Moro, A., On integrable conservation laws. 
Proceedings of the Royal Society of London A: Mathematical, Physical and Engineering Sciences, {\bf 471} (2015), No. 2173.

\bibitem{Buryak} 
Buryak, A., Dubrovin-Zhang hierarchy for the Hodge integrals. 
Commun. Number Theory Phys.,~{\bf 9} (2015), 239--272.

\bibitem{BPS1} Buryak, A., Posthuma, H., Shadrin, S., A polynomial bracket for the Dubrovin--Zhang hierarchies. 
Journal of Differential Geometry, {\bf 92} (2012), 153--185.

\bibitem{BPS2} Buryak, A., Posthuma, H., Shadrin, S.,  
On deformations of quasi-Miura transformations and the Dubrovin--Zhang bracket. Journal of Geometry and Physics, 
{\bf 62} (2012), 1639--1651.

\bibitem{BR24}
Buryak, A., Rossi, P., Counting meromorphic differentials on $\mathbb{C}\mathbb{P}^1$.
Lett. Math. Phys.,~{\bf 114} (2024), Paper No. 97, 27 pp.

\bibitem{DW}
Dijkgraaf, R., Witten, E., 
Mean field theory, topological field theory, and multi-matrix models. Nucl. Phys.~B,~{\bf 342} (1990), 486--522.

\bibitem{Du0}
Dubrovin, B.,  Integrable systems in topological field theory. Nucl. Phys. B, \textbf{379} (1992), 627--689.

\bibitem{Du1} 
Dubrovin, B.,  Geometry of 2D topological field theories. In ``Integrable Systems and
Quantum Groups" (Montecatini Terme, 1993), pp.~120--348, Francaviglia, M., Greco S. (Eds.), Springer Lecture Notes in Math., \textbf{1620}, 1996.

\bibitem{Du2} 
Dubrovin, B.,  Painlev\'e transcendents in two-dimensional topological field theory.
In ``The Painlev\'e Property: One Century Later", pp. 287--412, Conte, R. (Ed.). CRM Ser. Math. Phys., Springer, New York, 1999.

\bibitem{Du3} Dubrovin, B., 
On almost duality for Frobenius manifolds. 
In ``Geometry, topology, and mathematical physics", pp. 75--132. 
Amer. Math. Soc. Transl. Ser.~{\bf 2}, {\bf 212}, Amer. Math. Soc., Providence, RI, 2004.

\bibitem{DLYZ}
Dubrovin, B., Liu, S.-Q., Yang, D., Zhang, Y., 
Hodge integrals and tau-symmetric integrable hierarchy of Hamiltonian evolutionary PDEs. 
Adv. Math.,~{\bf 293} (2014), 382--435.

\bibitem{DVY}
Dubrovin, B., Valeri, D., Yang, D., 
Affine Kac--Moody algebras and tau-functions for the Drinfeld--Sokolov hierarchies: the matrix-resolvent method.
SIGMA,~{\bf 18} (2022), Paper No. 077, 32 pp.

\bibitem{DZ-oneloop}
Dubrovin, B., Zhang, Y., 
Bi-Hamiltonian hierarchies in $2D$
topological field theory at one-loop approximation. Comm. Math.
Phys.,~\textbf{198} (1998), 311--361.

\bibitem{DZ-vira}
Dubrovin, B., Zhang, Y., Frobenius manifolds and Virasoro
constraints. Selecta Math.,~\textbf{5} (1999), 423--466.

\bibitem{DZ-norm}  
Dubrovin, B., Zhang, Y., Normal forms of hierarchies of integrable PDEs, 
Frobenius manifolds and Gromov - Witten invariants. 
a new 2005 version of arXiv:math/0108160v1, 295~pp.

\bibitem{DZ-toda}
Dubrovin, B., Zhang, Y., Virasoro symmetries of the extended Toda hierarchy. Comm. Math. Phys., \textbf{250} (2004), 161--193.

\bibitem{EHX}
Eguchi, T., Hori K., Xiong, C.-S.,
Quantum Cohomology and Virasoro Algebra. Phys. Lett. B,~{\bf 402} (1997), 71--80.

\bibitem{FP} 
Faber, C., Pandharipande, R., 
Hodge integrals and Gromov-Witten theory. Invent. Math.,~\textbf{139} (2000), 173--199.

\bibitem{Getzler}
Getzler, E.,
Intersection theory on $\overline{\mathcal{M}}_{1,4}$ and elliptic Gromov-Witten invariants.
J. Amer. Math. Soc.,~{\bf 10} (1997), 973--998.

\bibitem{G1}
Givental, A., Gromov--Witten invariants and quantization of
quadratic Hamiltonians. Mosc. Math. J.,~\textbf{1} (2001), 551--568.

\bibitem{LWZ}
Liu, S.-Q., Wu, C.-Z., Zhang, Y., On properties of Hamiltonian structures for a class of evolutionary PDEs.
Lett. Math. Phys.,~{\bf 84} (2008), 47--63.

\bibitem{LWZZ}
Liu, S.-Q., Wu, C.-Z., Zhang, Y., Zhou, X.,
Drinfeld-Sokolov hierarchies and diagram automorphisms of affine Kac-Moody algebras. 
Comm. Math. Phys.,~{\bf 375} (2020), 785--832.

\bibitem{LYZZ} 
Liu, S.-Q., Yang, D., Zhang, Y., Zhou, J.,  
The Virasoro-like algebra of a Frobenius manifold.
IMRN,~{\bf 2023} (2023), 13524--13561.

\bibitem{LZ} 
Liu, S.-Q., Zhang, Y., On quasi-triviality and integrability of a class of scalar evolutionary PDEs. 
J. Geom. Phys.,~\textbf{57} (2006), 101--119.

\bibitem{LT} 
Liu, X., Tian, G., Virasoro constraints for quantum cohomology. 
Journal of Differential Geometry,~\textbf{50} (1998), 537--590.

\bibitem{Shima}
Shima, H., The geometry of Hessian structures.
World Scientific Publishing Co. Pte. Ltd., Hackensack, NJ, 2007, 246 pp.

\bibitem{Totaro}
Totaro, B., The curvature of a Hessian metric.
Internat. J. Math.,~{\bf 15} (2004), 369--391.

\bibitem{Witten} 
Witten, E., Two-dimensional gravity and intersection theory on moduli space.
Surveys in Differential Geometry,~{\bf 1} (1991), 243--310.

\bibitem{YZ}
Yang, D., Zagier, D., Mapping partition functions. arXiv:2308.03568.

\end{thebibliography}
\end{document}